# The dependence of scattering length on van der Waals interaction and on the reduced-mass of the system in two-atomic collision at cold energies

**Hasi Ray**[1,2,3,4*]

[1]Study Center, S-1/407/6, B. P. Township, Kolkata 700094, India
[2]Department of Physics, New Alipore College, New Alipore, Kolkata 700053, India
[3]Department of Science, NITTTR-K, Salt Lake, Kolkata 700106, India
[4]Code 671, Heliophysics Division, GSFC, NASA, Greenbelt, Maryland 20771, USA

[*]hasi_ray@yahoo.com

**Abstract:** The static-exchange model (SEM) and the modified static-exchange model (MSEM) recently introduced by Ray [1] is applied to study the elastic collision between two hydrogen-like atoms when both are in ground states considering the system as a four-body Coulomb problem in the center of mass frame, in which all the Coulomb interaction terms in direct and exchange channels are treated exactly. The SEM includes the non-adiabatic short-range effect due to electron-exchange. The MSEM added in it, the long-range effect due to induced dynamic dipole polarizabilities between the atoms e.g. the Van der Waals interaction. Applying the SEM code in different H-like two-atomic systems, a reduced mass ($\mu$) dependence on scattering length is observed. Again applying the MSEM code on H(1s)-H(1s) elastic scattering and varying the minimum values of interatomic distance $R_0$, the dependence of scattering length on the effective interatomic potential consistent with the existing physics are observed. Both these basic findings in low and cold energy atomic collision physics are quite useful and are being reported for the first time.

The elastic collision between two hydrogen-like atoms is the simplest and the most efficient tool to find the basic physics. In the present communication, the H(1s)-H(1s) elastic scattering is studied at low or cold energies to derive the s-wave elastic phase shift and the scattering length using the recently introduced [1] static-exchange model (SEM) and the modified static-exchange model (MSEM). The SEM includes the non-adiabatic short-range effect due to electron-exchange that is repulsive in triplet (-) channel. The MSEM includes the long-range effect due to van der Waals interaction in addition to the short-range effect introduced by the SEM. The center of mass frame is used to describe the four-body Coulomb system. All the Coulomb interaction terms in direct and exchange channels are calculated exactly to find the Born-Oppenheimer (BO) matrix elements [2] that acts as input to find the unknown SEM amplitude, following a coupled-channel methodology introduced by Calcutta Group [3]. The detailed of the theory is available in the literature [1-9]. Here Lippman-Schwinger type coupled integral equation in momentum space formalism [3] is used. The formally exact Lippman-Schwinger type coupled integral equation for the scattering amplitude in momentum space is given by [3]:

$$f^{\pm}_{n'1s,n1s}(\vec{k}_f,\vec{k}_i) = B^{\pm}_{n'1s,n1s}(\vec{k}_f,\vec{k}_i) - \frac{1}{2\pi^2}\sum_{n''}\int d\vec{k}'' \frac{B^{\pm}_{n'1s,n''1s}(\vec{k}_f,\vec{k}'')f^{\pm}_{n''1s,n1s}(\vec{k}'',\vec{k}_i)}{\vec{k}^2_{n''1s} - \vec{k}''^2 + i\varepsilon} \quad (1),$$

where $B^{\pm}$ are the well known Born-Oppenheimer (BO) scattering amplitude [2] in the singlet (+) and triplet (-) channels respectively. Accordingly $f^{\pm}$ indicate the unknown SEM scattering amplitudes for the singlet and triplet states of the two system electrons.

The present interest lays on triplet channel to find the basic physics of cold-atomic system. The BO amplitude for the triplet channel is defined as

$$B^{-}_{n'1s,n1s}(\vec{k}_f,\vec{k}_i) = -\frac{\mu}{2\pi}\int d\vec{R}d\vec{r}_1 d\vec{r}_2 \psi^*_f(\vec{R},\vec{r}_1,\vec{r}_2)V(\vec{R},\vec{r}_1,\vec{r}_2)\psi_i(\vec{R},\vec{r}_1,\vec{r}_2) \quad (3),$$

when $V(\vec{R},\vec{r}_1,\vec{r}_2)$ represent the Coulomb interaction: $V_{Direct}$ is for the direct channel and $V_{Exchange}$ is for the exchange or rearrangement channel; $\mu$ is the reduced mass of the system. $\vec{R}$ is the inter-nuclear separation and $\vec{r}_1$, $\vec{r}_2$ are the coordinates of two electrons from their corresponding nuclei with charges $+Z_1$ and $+Z_2$ in a.u. so that

$$V_{Direct}(\vec{R},\vec{r}_1,\vec{r}_2) = \frac{Z_1 Z_2}{R} - \frac{Z_1}{|\vec{R}-\vec{r}_2|} - \frac{Z_2}{|\vec{R}+\vec{r}_1|} + \frac{1}{|\vec{R}+\vec{r}_1-\vec{r}_2|} \quad (4),$$

$$V_{Exchange}(\vec{R},\vec{r}_1,\vec{r}_2) = \frac{Z_1 Z_2}{R} - \frac{Z_1}{|\vec{r}_1|} - \frac{Z_2}{|\vec{r}_2|} + \frac{1}{|\vec{R}+\vec{r}_1-\vec{r}_2|} \quad (5).$$

The initial and final state wavefunctions are defined respectively as

$$\psi_i = e^{i\vec{k}_i\cdot\vec{R}'}\phi_{1s}(r_1)\eta_{1s}(r_2) \quad (6),$$

$$\psi_f = (1-P_{12})e^{-i\vec{k}_f\cdot\vec{R}'}\phi_{1s}(r_1)\eta_{1s}(r_2) \quad (7);$$

$\phi_{1s}(r_1)$ and $\eta_{1s}(r_2)$ are two atomic wavefunctions and $P_{12}$ is the exchange or antisymmetry operator. $\vec{R}'$ represent the CM-coordinates of the system in the direct channel so that

$$\vec{R}' = \vec{R} + \frac{m_e}{m_A+m_e}\vec{r}_1 - \frac{m_e}{m_B+m_e}\vec{r}_2 \quad (8);$$

when $m_A$, $m_B$ are the masses of two positrons and $m_e$ is the electron mass. The CM-coordinates of the system in the rearrangement channel is defined by $\vec{R}_f$ and can be expressed as

$$\vec{R}_f = \vec{R} + \frac{m_e}{m_A + m_e}(\vec{r}_2 - \vec{R}) - \frac{m_e}{m_B + m_e}(\vec{r}_1 + \vec{R}) \tag{9}$$

The long-range matrix element due to van der Waals interaction that is added to BO amplitude in MSEM theory is

$$B_{van} = -\frac{\mu}{2\pi}\int d\vec{r}_1 \int d\vec{r}_2 \int d\hat{R} \int_{R=R_0}^{\infty} dR.R^2 \left[ \psi_f^*(\vec{R},\vec{r}_1,\vec{r}_2)\{-\frac{C_W}{R^6}\}\psi_i(\vec{R},\vec{r}_1,\vec{r}_2) \right] \tag{10}$$

when $C_W$ is the van der Waals coefficient [10] and $R_0$ is the minimum value of the interatomic distance. The s-wave elastic triplet phase shift and the corresponding scattering length are calculated following the standard procedure described in references [1-9].

The H-like two atomic systems of interests to study the reduced mass dependence of scattering lengths using the SEM code are the (i) Muonium (Mu) and Mu, (ii) Mu and H, (iii) Mu and Deuterium (D), (iv) Mu and Tritium (T), (v) H and H, (vi) H and D, (vii) H and T, (viii) D and D, (ix) D and T and (x) T and T. The reduced masses ($\mu$) of these systems in atomic units are respectively (i) $\mu = 103.9$, (ii) $\mu = 186.7$, (iii) $\mu = 196.7$, (iv) $\mu = 200.2$, (v) $\mu = 918.5$, (vi) $\mu = 1224.5$, (vii) $\mu = 1377.6$, (viii) $\mu = 1836.5$, (ix) $\mu = 2203.7$, (x) $\mu = 2754.5$. To vary the strength of van der Waals interaction in the MSEM code for H(1s)-H(1s) elastic scattering, the values of $R_0$ are varied from $2a_0$ to $20a_0$ to study the s-wave elastic phase shifts. Since two atoms cannot occupy the same space at a time, the minimum value of interatomic distance is chosen as $2a_0$.

Both the codes [1] are used to calculate the triplet (-) s-wave elastic phase shift ($\delta_0^-$) and corresponding cross section in the energy region with $k$ = .1, .2, .3, .4, .5, .6, .7, .8, .9 a.u. The variation of $k\cot\delta_0^-$ vs $k^2$ following the effective range theory for the values of $R_0 = 2a_0$, $3a_0$, $4a_0$, $5a_0$, $6a_0$, $7a_0$, $8a_0$, $9a_0$, $10a_0$, $11a_0$, $12a_0$, $15a_0$, $20a_0$ are presented in **Figure 1**. All the MSEM curves are made in solid lines; the bottom one is the MSEM data using $R_0 = 2a_0$ and the top solid curve is for $R_0 = 20a_0$. The dotted line is used to represent the SEM data [5]. All the curves are gradually laying upwards systematically as the value of $R_0$ gradually increased from $2a_0$ to $20a_0$; at $R_0 = 20a_0$, the MSEM curve almost coincide with the SEM curve (dotted line) indicating almost no contribution of long-range effect. The triplet scattering lengths corresponding to different values of $R_0$ are presented in **Table 1** and compared with the SEM data and other available data [11-15]. The MSEM data are improving gradually towards the reported scattering lengths as the strength of van der Waals interaction is increased gradually decreasing the value of $R_0$. The findings are consistent with the basic physics since the stronger attractive potentials cause the shorter scattering lengths and the stronger repulsive potentials cause the longer scattering lengths [16]. The long-range van der Waals interaction is always attractive. In addition, the larger reduced masses of the system make the interaction stronger [16].

The $k\cot\delta_0^-$ vs $k^2$ plot for H-like two-atomic systems with different reduced masses is presented in **Figure 2**. All the curves are made in solid lines: the bottom one is for the Mu-Mu system with reduced mass $\mu = 103.9$ a.u. and top one for T-T system with the reduced mass $\mu = 2754.5$ a.u. All the other curves lay gradually above the other with the gradual increase of reduced masses. An interesting and systematic resemblance are observed between the reduced mass and the scattering length of the system. The variation of the calculated scattering length with the corresponding reduced mass of the system are presented in **Table 2**. In Ps-Ps system $\mu = 1$ and the triplet scattering length is 3.25 a.u. using the present code [6]. The findings are consistent with the basic physics [16] because the static-exchange potential in triplet channel is repulsive and the larger reduced mass makes the interaction stronger. So accordingly the scattering length is gradually increasing with the increase of reduced mass of the system.

It should be noted again that the present SEM theory includes only the non-adiabatic short-range effects and there is no contribution of long-range interaction in it. But at low and cold energies, the long-range van der Waals interaction has an important role in addition to short-range non-adiabatic effects. So inclusion of both these effects are extremely useful to describe the cold energy atomic collision processes. It could provide better informations if it becomes possible to include the van der Waals interaction in all the systems with different reduced masses. However in the literature no data are available for the van der Waals coefficients to study the H-like two-atomic systems described here. The data using the appropriate values of the van der Waals coefficients for different two-atomic systems in MSEM code could explain more accurately the unexplained findings like the electron-like behaviour of Ps [17], the occurrence of Bose-Einstein condensation (BEC) [18] in $Rb_{85}$ - $Rb_{85}$ system with $\mu = 78048.5$ a.u. It should be noted that in electron-atom and positronium-atom collision, the system reduced masses are almost equal; as the atoms become heavier, the two reduced masses exactly coincide. Again the reduced masses of $Rb_{87}$ - $Rb_{87}$ system is $\mu = 79884.5$ a.u. and for $Rb_{85}$ - $Rb_{87}$ system is $\mu = 78955.8$ a.u., both are greater than the reduced mass of $Rb_{85}$ - $Rb_{85}$ system. It would be extremely useful to include the first excited 2s and 2p states of both the atoms in the coupled-channel methodology for more accurate data; indeed it is very difficult and extremely labourious job. On the other way, it would be useful to introduce the first excited 2s state of both the atoms following the coupled-channel methodology in the MSEM code to include the majority of the non-adiabatic short-range effects.

In conclusion we report for the first time the reduced mass dependence of scattering length in two-atomic collision at low and cold energies. Using the SEM theory in H-like two-atomic systems with reduced masses laying in the range 103.9 to 2754.4 a.u., it is observed that with the increase of reduced mass of the system, the scattering length increases almost proportionately. Again applying the MSEM theory in H(1s)-H(1s) elastic scattering, the variation of scattering length with the strength of attractive van der Waals interaction that is consistent with the existing physics, is observed. The most accurate calculation using the methods described above could provide more accurate informations.

The author would be glad to acknowledge the DST's support through grant no. SR/WOSA/PS-13/2009, Government of India to carry on the investigation if the grant is extended after November 30[th] 2012 till date when investigations are carried on. The author is thankful for the close association and academic discussions of A. K. Bhatia, A. Temkin, R. J. Drachman, J. D. Rienzi and others at GSFC NASA when revised the paper during her visit since July to October 2015.

**References**
1. H. Ray, Pramana J. Phys. **83** 907 (2014).
2. H. Ray and A. S. Ghosh, J. Phys. B **29** 5505 (1996); ibid **30** 3745 (1997).
3. A. S. Ghosh, N. C. Sil and p. Mandal, Phys. Rep. **87** 313-406 (1982).
4. The effect of van der Waals interaction in elastic collision between Ps(1s) and H(1s), H. Ray et al, arXiv: 1308.1939 (2013).
5. The collision between two hydrogen atoms, H. Ray, arXiv: 1311.3132 (2013).
6. The collision between two ortho-Positroium (Ps) atoms: a four-body Coulomb problem, H. Ray, arXiv:1409.3371 (2014); Pramana J. Phys. (IAS) **in press** (2015)..
7. The collision between positronium (Ps) and muonium (Mu), H. Ray, arXiv: 1501.00260 (2015); J. Phys. Conference Series (IOP Publishing) **618** 012008 (2015).

8. The atom and atom scattering model to control ultracold collisions, H. Ray, arXiv: 1206.5476 (2012).
9. The effect of long-range forces on cold-atomic interaction: Ps-H system, H. Ray, arXiv: 1103.4915 (2011).
10. J. Mitroy and M. W. J. Bromley, Phys. Rev. A **68** 035201 (2003).
11. A. Sen, S. Chakraborty and A. S. Ghosh, Europhys. Lett. **76** 582 (2006).
12. M. J. Jamieson, A. Dalgarno and J. N. Yukich, Phys. Rev. A **46** 6956 (1992).
13. M. J. Jamieson and A. Dalgarno, J. Phys. B **31** L219 (1998).
14. C. J. Williams and P. S. Julienne, Phys. Rev. A **47** 1524 (1995).
15. N. Koyama and J. C. Baird, J. Phys. Soc.Jpn. **55** 801 (1986).
16. R. J. Drachman, Private communication during August-September 2015 at GSFC, NASA, Greenbelt, Maryland, USA.
17. S. J. Brawley, S. Amritage, J. Beale, D. E. Leslie, A. I. Williams and G. Laricchia, Science **330** 789 (2010).
18. S. Chaudhuri, S. Roy and C. S. Unnikrishnan, Current Science **95** 1026 (2008).

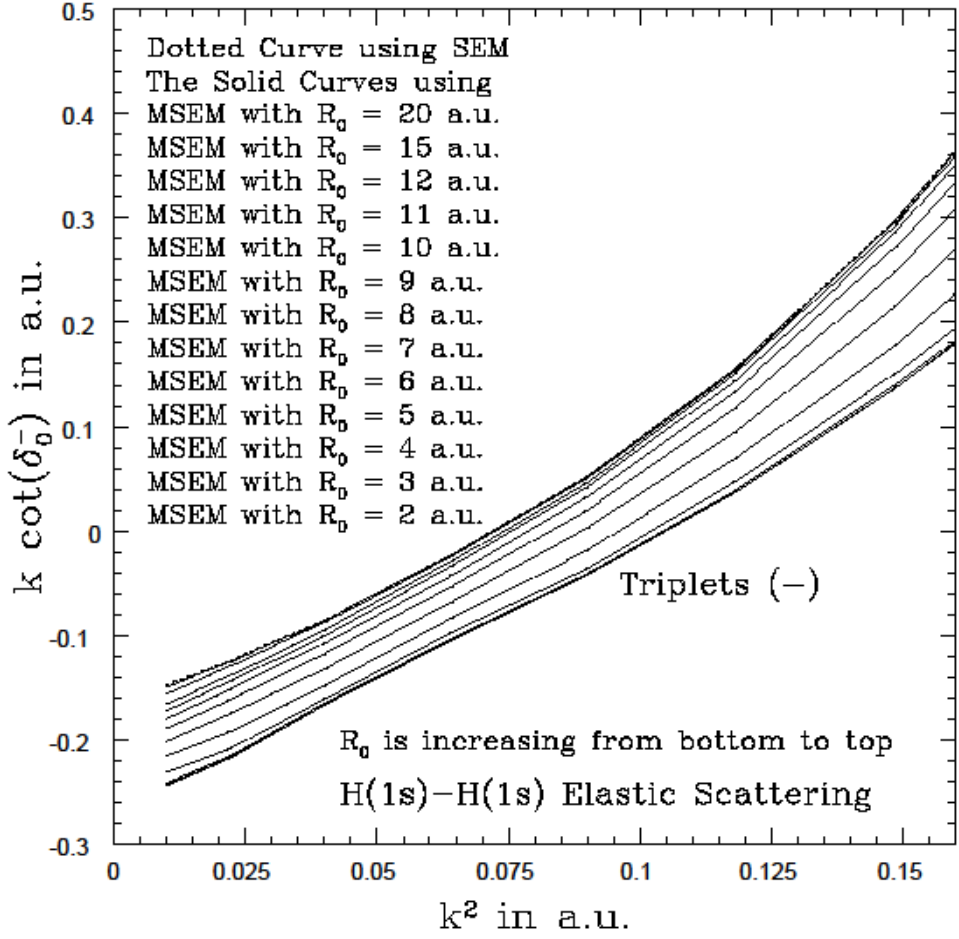

**Figure 1.** The $k\cot\delta_0^-$ vs $k^2$ plot for H(1s)-H(1s) elastic scattering using SEM and MSEM code for $R_0 = 2a_0$, $3a_0$, $4a_0$, $5a_0$, $6a_0$, $7a_0$, $8a_0$, $9a_0$, $10a_0$, $11a_0$, $12a_0$, $15a_0$, $20a_0$.

**Table 1.** The scattering length in atomic units using SEM and MSEM for different values of $R_0$ in H(1s)-H(1s) elastic scattering.

| Scattering length in atomic unit (a.u.) | | | | | | | | | | | | | | |
|---|---|---|---|---|---|---|---|---|---|---|---|---|---|---|
| Using SEM | Using MSEM with $R_0 =$ | | | | | | | | | | | | | Data of others |
| | $20a_0$ | $15a_0$ | $12a_0$ | $11a_0$ | $10a_0$ | $9a_0$ | $8a_0$ | $7a_0$ | $6a_0$ | $5a_0$ | $4a_0$ | $3a_0$ | $2a_0$ | |
| 5.88, 5.90[a] | 5.80 | 5.68 | 5.26 | 5.11 | 4.89 | 4.63 | 4.38 | 4.03 | 3.77 | 3.68 | 3.63 | 3.60 | 3.58 | 2.04[a], 1.91[b], 1.22[c], 1.34[d], 1.3[e] |

[a]Sen, Chakraborty and Ghosh [11];
[b]Jamieson, Dalgarno and Yukich [12];
[c]Jamieson and Dalgarno [13];
[d]Williams and Julienne [14];
[e]Koyama and Baird [15].

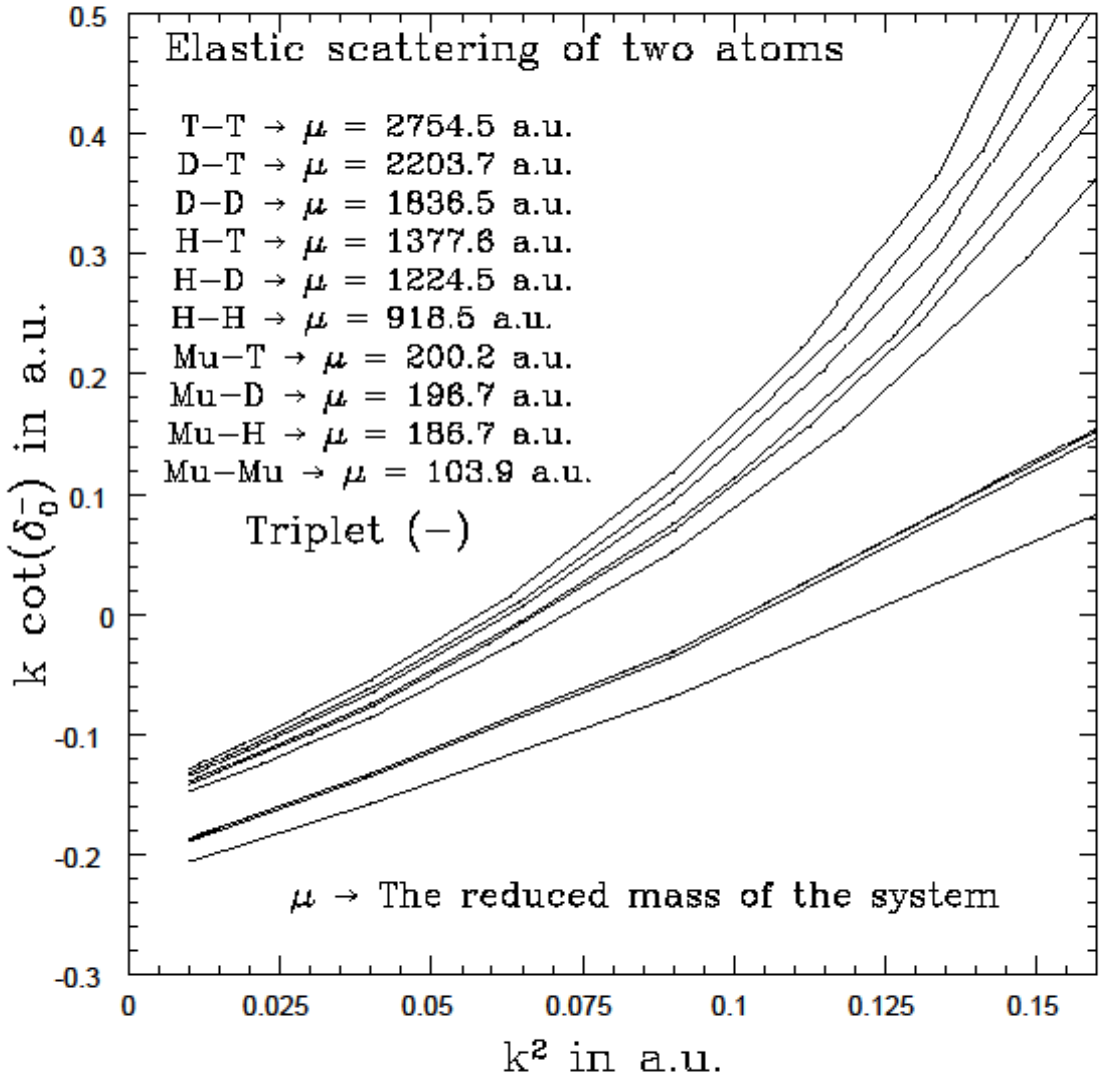

**Figure 2.** $k \cot \delta_0^-$ vs $k^2$ curve for different H-like two-atomic systems. The reduced masses of the systems are increasing gradually from bottom to top solid curves.

**Table 2.** The variation of triplet scattering length with reduced mass of different H-like two-atomic systems.

| System → | Mu-Mu | Mu-H | Mu-D | Mu-T | H-H | H-D | H-T | D-D | D-T | T-T |
|---|---|---|---|---|---|---|---|---|---|---|
| Reduced mass → in a.u. | 103.9 | 186.7 | 196.7 | 200.2 | 918.5 | 1224.5 | 1377.6 | 1836.5 | 2203.7 | 2754.5 |
| Scattering length → in a.u. | 4.54 | 4.76 | 4.88 | 4.95 | 5.88 | 6.25 | 6.37 | 6.58 | 6.68 | 6.90 |